# Dark Matter in the Central Region of NGC 3256

Israa Abdulqasim Mohammed Ali[1*], Chorng-Yuan Hwang[2], Zamri Zainal Abidin[1], Adele Laurie Plunkett[3]

[1]*Physics Department, University of Malaya, Kuala Lumpur, 50603, Malaysia.*

[2]*Institute of Astronomy, National Central University, Jhongli, 32001, Taiwan.*

[3]*European Southern Observatory (ESO), Alonso de Cordova 3107, Vitacura, Casilla 19001, Santiago, Chile*

[*]*Corresponding author: israa.aq@siswa.um.edu.my / israa.aq88@gmail.com*

## ABSTRACT

We investigated the central mass distribution of the luminous infrared galaxy NGC 3256 at a distance of 35 Mpc by using CO(1-0) observations of the Atacama Large Millimeter and sub-millimeter Array (ALMA) and near-IR data of the Two Micron Sky Survey (2MASS). We found that there is a huge amount of invisible dynamical mass ($4.84 \times 10^{10} M_\odot$) in the central region of the galaxy. The invisible mass is likely caused by some dark matter, which might have a cuspy dark matter profile. We note that this dark matter is difficult to explain with the conventional Modified Newtonian Dynamics (MOND) model, which is only applicable at a low acceleration regime, whereas the acceleration at the central region of the galaxy is relatively strong. Therefore, this discovery might pose a challenge to the conventional MOND models.

**Abstrak**

Kami telah menjalankan kajian terhadap taburan jisim di kawasan pusat galaksi inframerah terang NGC 3256 pada jarak 35 Mpc dengan menggunakan cerapan CO(1-0) dari Atacama Large Millimeter and sub-millimeter Array (ALMA) dan maklumat inframerah dekat dari Two Micron Sky Survey (2MASS). Penemuan kami menunjukkan terdapat jumlah jisim dinamik ghaib yang besar ( $4.48 \times 10^{10} M_\odot$ ) di kawasan pusat galaksi. Jisim ghaib ini berkemungkinan besar merupakan jirim gelap, yang mempunyai profil "cuspy". Hal tersebut sukar diterangkan dengan model "Modified Newtonian Dynamics" (MOND), yang hanya terpakai untuk kadar pecutan yang rendah, tetapi pecutan di kawasan pusat galaksi agak tinggi. Oleh yang demikian, penemuan ini mencabar model MOND konvensional.

**Kata kunci: jirim gelap---perubahan---galaksi (NGC 3256)**

**Introduction**

The cosmic microwave observations of Planck found that about 84% of mass in the universe is invisible (Ade et al. 2016). The existence of dark matter can be inferred from observations of gravitational influence on the movement of ordinary matter and also on electromagnetic wave, e.g., gravitational lensing. Observations of spiral galaxies suggested that a significant amount of invisible (missing) mass is required in order to explain the observed rotation curves of the spiral galaxies (Roberts & Whitehurst 1975; Faber & Gallagher 1979); the rotation curves of spiral galaxies have also been studied using different methods (Rubin et al. 1980; Bosma 1981). Besides, it was also found that different types of galaxies might have different distributions of dark matter; for example, dwarf galaxies might contain larger fractions of dark matter than normal galaxies (Faber & Lin 1983; Mateo 1998). All these studies showed that dark matter is a common feature in galaxies. However, although dark matter might contribute about 80% of the material mass of the universe (Bauer et al. 2015), attempts to detect the dark matter directly are as yet still unsuccessful (Tan et al. 2016).

An alternative theory, Modified Newtonian Dynamics (MOND)(Milgrom 1983), has been proposed to explain the observed properties of galaxies without utilizing dark matter. The theory proposed a modification of Newton's laws to account for the observed rotation curves of galaxies and suggested that the missing mass problem in galaxies only occurs in the small acceleration regime with $a \ll a0$, where $a0 \approx 1.2 \times 10^{-8}$ cm s$^{-2}$. The MOND theory can reproduce the galactic rotation curves from the observed distribution of stellar and gaseous matter (e.g. Bottema et al. (2002)).

The MOND theory could also account for the Tully– Fisher and Faber–Jackson relations of galaxies (van den Bosch & Dalcanton 2000) and the kinematics of small groups of galaxies (Milgrom 1998) without dark matter. Recently, Tian & Ko (2016) found that some elliptical galaxies seem to have a deficit of dark matter, and the dynamics of the elliptical galaxies can be well explained with the MOND theory.

Although MOND is quite successful in explaining many dark matter phenomena at the galactic scale, it is also well known that MOND theory cannot easily explain the phenomena of dark matter in galaxy clusters (e.g. Aguirre et al. 2001). It is thus very interesting to know whether the MOND theory at the galactic scale is really showing evidence for a modification of the Newtonian law, or it is only a coincidence that MOND has some similarity with dark matter. To investigate these possibilities, it is important to know whether MOND can really explain all dark matter phenomena at different galactic scales.

In the hierarchical scenario of dark-matter halo formation, i.e., galaxy formation, large halos (large galaxies) are formed from the merger of small halos (small galaxies). Luminous infrared galaxies (LIRGs), which were usually found to be in the process of galaxy merging, should thus also be in the processes of dark-matter halo, and would be a very important stage of galaxy evolution. However, LIRGs were usually found to contain a huge amount of molecular gas ($\geq 10^9 M_\odot$), which was usually too massive to be reconciled with the observed dynamical masses (e.g., Solomon et al. 1997). The inconsistency is usually believed to be caused by the CO-luminosity-to-H$_2$-mass conversion factor, which was derived from the galaxy and might not be applicable to the LIRGs. Even with adopting a lower value of conversion factor, the dynamical mass would still be dominated by the molecular mass. However, since LIRGs are usually merging galaxies, which are regulated by dark-matter halo merging, it would be interesting to ask whether we can observe a significant amount of dark matter in the centre regions of LIRGs, whose dynamical masses usually seem to be dominated by molecular masses.

The LIRG NGC 3256 is among the brightest nearby galaxies. It belongs to the Hydra-Centaurus clusters of galaxies at a distance D = 35 Mpc (Sargent et al. 1989). This galaxy is descended from a merger of two gas-rich disk galaxies and has a nearly face-on orientation (Agüero & Lipari 1991). These facts make the galaxy the best target to study the dark matter in a LIRG. NGC 3256 has two nuclei with 5″ (0.8 kpc) separation in the north-south direction (Alonso-Herrero et al. 2006). The southern nucleus is indeed heavily obscured; various studies argued that the southern nucleus may harbor a highly obscured AGN (Neff et al. 2003). It was also suggested that IR emission of these two nuclei are ascribed mostly to star formation process (Alonso-Herrero et al. 2012).

The molecular gas distributions of NGC 3256 have been variously studied (Sargent et al. 1989; Aalto et al. 1991; Baan et al. 2008). Submillimeter Array (SMA) CO(2-1) observations showed that there is a large molecular disk concentrated around the centre of the double nuclei using CO(2-1) (Sakamoto et al. 2006). Sakamoto et al. (2006) also discovered a high velocity component (up to around 400 km s$^{-1}$) of molecular gas at the center between the double nuclei. Sakamoto et al. (2014) used the Atacama Large Millimeter/submillimeter Array (ALMA) to study NGC 3256, and suggested that the two nuclei have their own molecular disks, and also that high-velocity gas is associated with the molecular outflows from the two nuclei and could have a velocity of ≥ 1000 km s$^{-1}$. In the present work, we determined the mass distributions of NGC 3256 using data from the Atacama Large

Millimeter and sub-millimeter Array (ALMA) and the Two Micron All Sky Survey (2MASS). The data reduction is described in section 2. Section 3 presents our results and discussion. Conclusions of our results are summarized in section 4.

**Data Selection**

The ALMA data of NGC 3256 were obtained from the ALMA Cycle 0 observations of NGC 3256 at the CO(J=1-0) rotational transition line (ID =2011.0.00525.S, PI: Sakamoto). The observations were carried out in 2011– 2012 using up to 23 12-m antennas. The central position of the observations was R.A. = 10h 27m51.23s, $Dec. = -43°54'16.6''(J2000)$.

The ALMA data were calibrated from raw data using the Common Astronomy Software Application (CASA) reduction package version 4.1. Data reduction and imaging processing were also performed with CASA. The CO (1-0) data cube was obtained from the continuum subtracted visibilities using the task 'uvconsub'. The cleaned images were obtained using the multiscale clean with natural weighting. The resulting beam size of the images is $\approx 2.3'' \times 2.10''$, corresponding to the angular linear scales of $390 pc \times 356 pc$. Figures 1 and 2 show the channel maps with velocity channel width= 20 km s$^{-1}$, and the integrated intensity map of the CO line in the central region of NGC 3256.

We used the 2MASS image of NGC 3256 to estimate the stellar mass of this galaxy. The data are publicly available. The 2MASS project is a ground-based near-infrared survey that observed the entire sky at the $J(1.2\mu)$, H ($1.65\mu$) and Ks ($2.17\mu$)bands (Finlator et al. 2000; Skrutskie et al. 2006; Schombert & Smith 2012). To estimate the stellar mass, we considered the Ks image of NGC 3256 from the 2MASS All-Sky Extended Source Catalog (XSC). Figure 3 shows the 2MASS Ks image in the centre region of NGC 3256.

## Analysis, Results and Discussion

### Central rotation curve

We examined the dynamic mass in the central region of NGC 3256 using the rotation curve of the galaxy derived from the CO line observations. To determine the rotation curve of NGC 3256, we first considered the position-velocity (PV) diagrams across the north nucleus (N nucleus), south nucleus (S nucleus) and the dynamic center of the galaxy along different positions angles with a slit of 5″ width.

We adopted the position for the north nucleus at R.A. (N)=10h27m51.23s, Dec.(N)= -43°54′14.0″ and that for the south nucleus at R.A.(S)=10h27m51.22s, Dec.(S)=-43°54′ 19.2″ (Neff et al. 2003). The center position between the two nuclei was taken to be R.A.=10h27m51.23s, Dec.= -43°54′16.6″ in J2000.0 (Lira et al. 2002; Sakamoto et al. 2006) (See Table 1). Figure 4 shows examples of the PV diagrams.

We note that the two nuclei of NGC 3256 are separated in the north-south direction by 5″ (0.8 kpc). Therefore, the rotation velocity of the gas should be less affected by the two nuclei along the direction of PA= -90°. We considered the velocity difference at PA≈ -90° to be caused mainly by the rotational velocity of the whole molecular distribution. The radial component of the rotation velocity on the line of sight $v_r$ is related to the observed radial velocity $v_{obs}$ by:

$$v_r = v_{obs} - v_{sys} , \qquad (1)$$

where, $v$sys is the systemic velocity of the galaxy. We assumed $v$sys to be 2775 km s$^{-1}$ from previous studies for NGC 3256 (Roy et al. 2005). The rotational velocity $v_c$ can be obtained as $v_c = \frac{v_r}{sin(i)}$. Figure 5 shows the projected rotation velocity along the line of sight of the system centered at R.A.=10h27m51.23s and Dec.=- 43°54″16.6″ from the PV diagram of Figure 4c.

## Dynamical and Molecular Gas Masses

The dynamical mass inside a radius r of a galaxy can be estimated with the rotation velocity $v_c$ at r (Koda et al. 2002).

$$M_{dyn} = 2.3 \times 10^5 \times \left(\frac{r}{kpc}\right) \times \left(\frac{v_c}{kms^{-1}}\right)^2 M_\odot, \quad (2)$$

We took the central position of the system to be R.A.=10h27m51.23s and Dec.=- 43°54′16.6″ to estimate the dynamical mass within the central region. From Figure 5, the $M_{dyn}$ is estimated to be $5.5\pm0.12\times10^{10}$ $M_\odot$ within r = 10″ from the line-of-sight rotational curve of ~ 187±4.30 km s$^{-1}$ assuming the inclination angle i = 30° from (Sakamoto et al. 2014). Based on Hughes & Hase (2010), we estimated the uncertainty of $M_{dyn}$. We assumed the function (f= aA$^b$) for the total mass of NGC 3256, and the corresponding error is $\frac{fb\sigma_A}{A}$. A' here is the value of velocity, with the standard deviation of $\sigma_A$ and b is a constant, which is 2. On the other hand, if the molecular gas is not rotating but moving randomly, we can estimate the dynamical mass using the virial theorem. The velocity dispersion along the line of sight $\sigma$ is about 82 km s$^{-1}$ from a Gaussian fit at the central region of NGC 3256. The total mass can be obtained from $M_{tot}$= 5v$^2$r/$\alpha$G according to McKee & Zweibel 1992, with $\alpha$ = 1 – 2, and $v^2$= $3\sigma^2$. We thus have $M_{tot} \sim (2 - 4 \times 10^{10}$ $M_\odot)$. The mass is close to the mass derived using the rotational curve.

The molecular hydrogen mass of NGC 3256 can be estimated using (Tsai et al. 2009):

$$M_{H2} = 1.2 \times 10^4 \times D^2 \times S_{CO(1-0)} \times \frac{X_{CO}}{3\times10^{20}}, \quad (3)$$

where, D is the distance in units of Mpc, $S_{CO(1-0)}$ is the CO-line flux in units of Jy km s$^{-1}$ and $X_{CO}$ is the CO-to-$H_2$ conversion factor. The conversion factor is usually in the range of $X_{CO}$=

$0.3 \times 10^{20}$ - $3 \times 10^{20}$ cm$^{-2}$ (K km s$^{-1}$)$^{-1}$ with high values for normal galaxies (see (Dickman et al. 1986)), and low values for LIRGs (Bolatto et al. 2013). The total gas mass associated the molecular hydrogen is $M_{gas} \approx 1.36 \times M_{H2}$ (Tsai et al. 2009). The total CO (1-0) flux is $\approx 813 \times 16$ Jy km s$^{-1}$ within the radius 10″ of NGC 2356; this region contains almost all of the CO(1-0) emission. We assumed the $X_{CO}$ to be $0.3 \times 10^{20}$ cm$^{-1}$ (K km s$^{-1}$)$^{-1}$ and found $M_{H2} = 1.19 \pm 0.23 \times 10^{9}$ M$_\odot$ and $M_{gas} \approx 1.62 \pm 0.32 \times 10^{9}$ M$_\odot$ within the central region. Our results for the molecular mass are also consistent with those derived by Sakamoto et al. (2014). To estimate the uncertainty in the molecular hydrogen mass and total gas mass of NGC 3256, we assumed that f=aA (Hughes & Hase (2010)). 'A' here is the mass and 'a' is the constant. The uncertainty of this function is /a/ $\sigma_A$, where $\sigma_A$ is the error of the integrated flux density.

Using the SEST 15 m single dish observation, Casoli et al. (1991) found that the $M_{H2}$ within the area 44″× 44″ is equal to $1.36 \times 10^{10}$ M$_\odot$ at D =35 Mpc, which is close to our result of $M_{H2} = 1.41 \times 10^{10}$ M$_\odot$ from ALMA observation at the same area. This result indicates that no missing flux exists.

**Stellar Mass**

The central region of NGC 3256 might also contain a significant amount of stellar mass. We used 2MASS Ks band images to derive the stellar mass of NGC 3256. To obtain the stellar mass in the central region of the galaxy, we need to de-project the image profile I(R) to obtain the luminosity density j(r) distribution of the galaxy and converted the luminosity density distribution to stellar mass. In general, if the image profile I(R) is circular symmetric, we can de-project the I(R) to find the j(r): $j(r) = \frac{-1}{\pi} \int_r^\infty \frac{dI}{dR} \frac{dR}{\sqrt{R^2-r^2}}$. We could derive j(r) numerically using the above formula. However, we found that the Ks image of NGC 3256 can be well fitted with reduced $\chi^2$ ($\chi^2_{red} \approx 1$) with the modified Hubble profile (Binney&Merrifield 1998): $I(r) = \frac{I_0}{1+\left(\frac{r}{r_0}\right)^2}$, which has a simple analytic form of j(r): $j(r) = \frac{j_0}{\left[1+\left(\frac{r}{r_0}\right)^2\right]^{\frac{3}{2}}}$, where $I_0$ is the central surface brightness, $r_0$ is the core radius, and $r_0$ and $j_0$ is related to the central surface brightness with $I_0 = 2r_0 j_0$. The de-projected total luminosity from the central region within a radius R can be derived as $L = \int_0^R 4\pi r^2 j(r) dr$; and the de-projected flux from the central region would be $F = L/(4\pi D^2)$: Figure 6 shows the brightness profile of the 2MASS Ks image of NGC 3256 and the fitted modified Hubble profile. The well fit of the Ks image with the modified Hubble profile makes it much easier to derive the luminosity density distribution of the galaxy. Besides, we also tried to fit the stellar light profile with the de Vaucouleurs $r^{1/4}$ law but was unable to obtain a reasonable fit ($\chi^2_{red} \approx 20$). We note that the modified Hubble profile is an approximation of the King model, which is derived by assuming that the stellar system is in a quasi-relaxation state (King 1966). The well fit of the Ks image with the modified Hubble profile thus suggests that the system should be very close to be dynamically relaxed. To obtain the stellar mass of NGC 3256, we first converted the de-projected flux within the selected galactic region to apparent magnitude:

$$m = 19.93 - 2.5 log(F);$$

where 19.93 is the zero point magnitude of the Ks band (Zhang et al. 2012). We then converted the apparent magnitude to absolute magnitude using (Kochanek et al. 2003) $M_K = m - 25 - 5log(DL) - k_z$, where, $D_L$ is the luminosity distance in Mpc, and $k_z$ = -6.0log(1+ z) is the k-correction (Domingue et al. 2009). The derived absolute magnitudes is about -20.45. We assumed that the absolute magnitude of the Sun at the Ks band is 3.39 (Mulroy et al. 2014), and we then found that the Ks luminosities are about $4.38 \times 10^9$ $L_\odot$ within r = 10″ of the central region of NGC 3256. The mass-to-light ratio at the 2MASS Ks band is about M/L =0.73 (Bell et al. 2003). The stellar mass in the central region of NGC 3256 is then estimated to be about $M_* = 3.06 \pm 0.27 \times 10^9 \, M_\odot$ within r = 10″. The uncertainty of the stellar mass estimated is the same as the molecular hydrogen mass and total gas mass of NGC 3256.

### Dark Matter in the Central Region of NGC 3256

The dynamical mass we derived from the rotation curve should include all baryonic masses, such as stars, gas, dust, and super massive black holes (SMBHs), as well as dark matter within the central region of NGC 3256:

$M_{dyn} = M_{DM} + M_{gas} + M_* + M_{HI} + M_{SMBH} + M_{dust}$.

The mass of the SMBH, dust, and HI gas can be estimated from previous observations with well-established empirical relations. The SMBH mass of NGC 3256 is expected to be only around $10^7 - 10^8 \, M_\odot$ based on the SMBH-bulge relation (Alonso-Herrero et al. 2013) and is around $1.7 \times 10^7 \, M_\odot$ estimated from the X-ray and radio luminosities (Merloni et al. 2003). The dust mass of NGC 3256 is around $5 \times 10^6 \, M_\odot$ estimated from the IRAS flux densities at $60\mu$ m and $100\mu$ m obtained from the NED and the relation derived by (Riffel et al. 2015). The dust mass and the SMBH mass are thus negligible compared with the molecular mass in the central region of NGC 3256. The HI observations of NGC 3256 shows absorption features in the central regions (English et al. 2003). The HI mass was estimated by (English et al. 2003) to be around $0.4 - 2 \times 10^9 \, M_\odot$ within the synthesized beam of their observation (23" in diameter). The baryonic mass within the 10" central region of NGC 3256 is thus about $M_{baryon} \approx M_{gas} + M_* + MHI \approx 6.62 \pm 0.40 \times 10^9 M_\odot$: Please note that most baryonic mass is from molecular mass and stellar mass; HI mass contributes very little. Comparing that with the dynamical mass we derived, we found that there is about $4.84 \pm 0.42 \times 10^{10} \, M_\odot$ invisible mass in the central region. The fraction of the invisible mass is about 87% of the dynamical mass. As mentioned earlier, the estimated uncertainty is based on error methods in Hughes &

Hase (2010). We note that Sakamoto et al. (2014) found the gas-to-dynamical mass ratio to be around 9%, a difference caused mainly by the fact that they adopted a different CO-to-$H_2$ conversion factor and they did not consider the stellar mass and HI mass. We would obtain similar results if we had adopted the same parameters. However, we used the most conservative parameters to make sure of the reality of the existence of invisible mass. In any case, a significant amount of invisible mass in the central region of the galaxy is unavoidable. The observed rotation velocity might have been influenced by the merging process of NGC 3256. However, since we used the velocity of the molecular gas to derive the dynamical mass, we would have overestimated the dynamical mass only if the molecular gas was unbounded to the galaxy. Some of the molecular outflows from the nuclei of NGC 3256 might be unbounded, but the outflow velocities are distinguishable from the rotation velocity and would not be confused (Sakamoto et al. 2014). We have avoided the regions of nuclei in deriving the rotation velocity. Therefore, the rotation velocity should not be affected by the molecular outflows. Random velocity of the molecular gas associated with stellar /AGN feedback may lead to some uncertainties in the estimate of the dynamical mass. However, the derived rotation velocity showed a smooth rotation curve with small fluctuations, indicating that the influence of the random velocity should be negligible.

It is clear from Figure 7 that the baryonic component in the central region of NGC 3256 is very much smaller than the total dynamical mass, reflecting the fact that the mass in the central region of the galaxy is dominated by dark matter. Additionally, the mass distribution of the baryonic component is very different from that of the dynamical one, suggesting the distribution of the baryonic mass and that of the dark matter is also different.

Most of the invisible mass should be in the form of dark matter. The amount of dark matter is about $4.84 \pm 0.42 \times 10^{10} M_\odot$, which is significantly larger than the stellar mass. It is clear that even the velocity was not circular, the dynamical mass that would be required to account for the observed velocity dispersion is still much larger than the baryonic mass, and thus the existence of a huge amount of dark matter in the central region of the galaxy is necessary. Dark matter is the dominant mass component for most galaxies and is the fundamental ingredient in determining the properties and evolution of the galaxies. However, dark matter usually dominates in the outer regions of galaxies but is not considered as an important mass component in the central region of galaxies. The dark matter in the innermost regions of galaxies is relatively unexplored. In a recent study, Iocco et al. (2015) investigated the dark matter in the Milky Way up to a radius larger than 2.5 kpc, using the motion of stars and HI gas. However, it is difficult to constrain MOND at r > 2.5 kpc. The discovery of a significant amount of dark matter in the central region of a galaxy might have a significant impact on MOND theories, which were proposed to explain the "missing mass problem" without using dark matter. A MOND theory should become the normal Newtonian dynamics at high acceleration, i.e., when the acceleration a is much larger than a constant $a_0$, which was found to be $\approx 1:2 \times 10^{-8}$ cm s$^{-2}$. On the other hand, at the low acceleration limit, where a $\ll a0$, the Newtonian force $F_N$ is related to the acceleration with the form (Sanders & Verheijen 1998; Bottema et al. 2002; Sanders & Noordermeer 2007):

$$F_N = m\mu\left(\frac{a}{a_o}\right)a, \tag{4}$$

where $F_N$ is the Newtonian force and $\mu$ (x) is the "interpolating function". The exact form of the interpolating function is yet to be determined, but it should have the approximations: $\mu$ (x) $\approx$ 1 for x $\gg$ 1 and $\mu$ (x) $\approx$ x for x $\ll$ 1. Two common choices for the interpolating functions have the forms:

$$\mu\left(\frac{a}{a0}\right) = \left[1 + \left(\frac{a0}{a}\right)\right]^{-1}, \tag{5}$$

And

$$\mu\left(\frac{a}{a0}\right) = \left[1 + \left(\frac{a0}{a}\right)^2\right]^{\frac{-1}{2}} \tag{6}$$

An implication of the MOND theory is that we would not expect to see the dark matter phenomenon in the region of high accelerations where the dynamics should be well described by the normal Newtonian law. However, we found that the acceleration in the central region of NGC 3256 is about a = $v^2/r \approx 2.67\times10^{-7}$ cm s$^{-2}$ at r = 1.7 kpc (see Figure 5). The derived $\mu$ (x) are$\approx$ 0.95 and 0.99 for both interpolating functions respectively. Both $\mu$ (x) are too large to account for the missing mass. Therefore, it might be impossible to explain the mass discrepancy in the centre region of NGC 3256 with the traditional MOND theory. However, it was well known that the standard $X_{CO}$, derived from molecular clouds in the Galactic plane, might over-estimate the molecular mass in the central regions of LIRGs (Hinz & Rieke 2006). It was suggested the true conversion factor in LIRGs might be a factor of 5 lower than the standard value. Since NGC 3256 is an LIRG, it might be more accurate to adopt a small conversion factor for the molecular gas in the central region of the galaxy. Sakamoto et al. (2014) argued that they did not know the true $X_{CO}$ in NGC 3256 and did not have a strong reason to believe that $X_{CO}$ is constant across the galaxy. Therefore Sakamoto used $X_{CO}$= $1\times10^{20}$ cm$^{-2}$ (K km s$^{-1}$)$^{-1}$ to derive the molecular mass of NGC 3256. We emphasize that different conversion factors do not affect our conclusion because we used $X_{CO}$ = $0.3\times10^{20}$ cm$^{-2}$ (K km s$^{-1}$)$^{-1}$, which is the smallest possible value of $X_{CO}$. If we used a different $X_{CO}$, $X_{CO}$=1 $\times10^{20}$cm$^{-2}$ (K km s$^{-1}$)$^{-1}$, the dark matter could be as massive as 82% of the dynamical mass. This also makes it difficult to account for the invisible mass with the MOND theory.

**Conclusions**

We reported the mass distributions of NGC 3256 using ALMA CO(1-0) observations and the NIR data of 2MASS. We showed that within the central region of the galaxy, there is a significant amount of invisible mass, which cannot be explained by the molecular mass and the stellar mass within this region. We suggested that there might be a significant amount of dark matter in the central region of the galaxy. What is more important is that this missing mass problem cannot be explained with the traditional MOND theory because of strong acceleration in the central region of the galaxy. Since most of the dark matter phenomena at galactic scales can usually be explained by the MOND theory very well, this discovery thus poses a significant challenge to the traditional MOND models. The study of Dark Matter is a significantly important research internationally as well as in Malaysia due to its novelty in research especially relating to radio astronomy and elementary particle physics. The first radio telescope in Malaysia (operated by the Radio Cosmology Research Lab) aims to study such topics and Malaysian researchers from the National Centre for Particle Physics have also been investigating dark matter (through their collaboration with the European Organization for Nuclear Research. or CERN).

## Acknowledgements

This paper makes use of the following ALMA data: ADS/JAO.AL MA#2011.0.00525.S. ALMA is a partnership of ESO (representing its member states), NSF (USA) and NINS (Japan), together with NRC (Canada), NSC and ASIAA (Taiwan), and KASI (Republic of Korea), in cooperation with the Republic of Chile. The Joint ALMA Observatory is operated by ESO, AUI/NRAO and NAOJ. The author Zamri Zainal Abidin would also like to acknowledge the University of Malaya's HIR grant UM.S/ 625/3/HIR/24 for their funding. CYH acknowledge support from the Ministry of Science and Technology (MOST) of Taiwan through grant MOST 103-2119-M-008- 017-MY3.

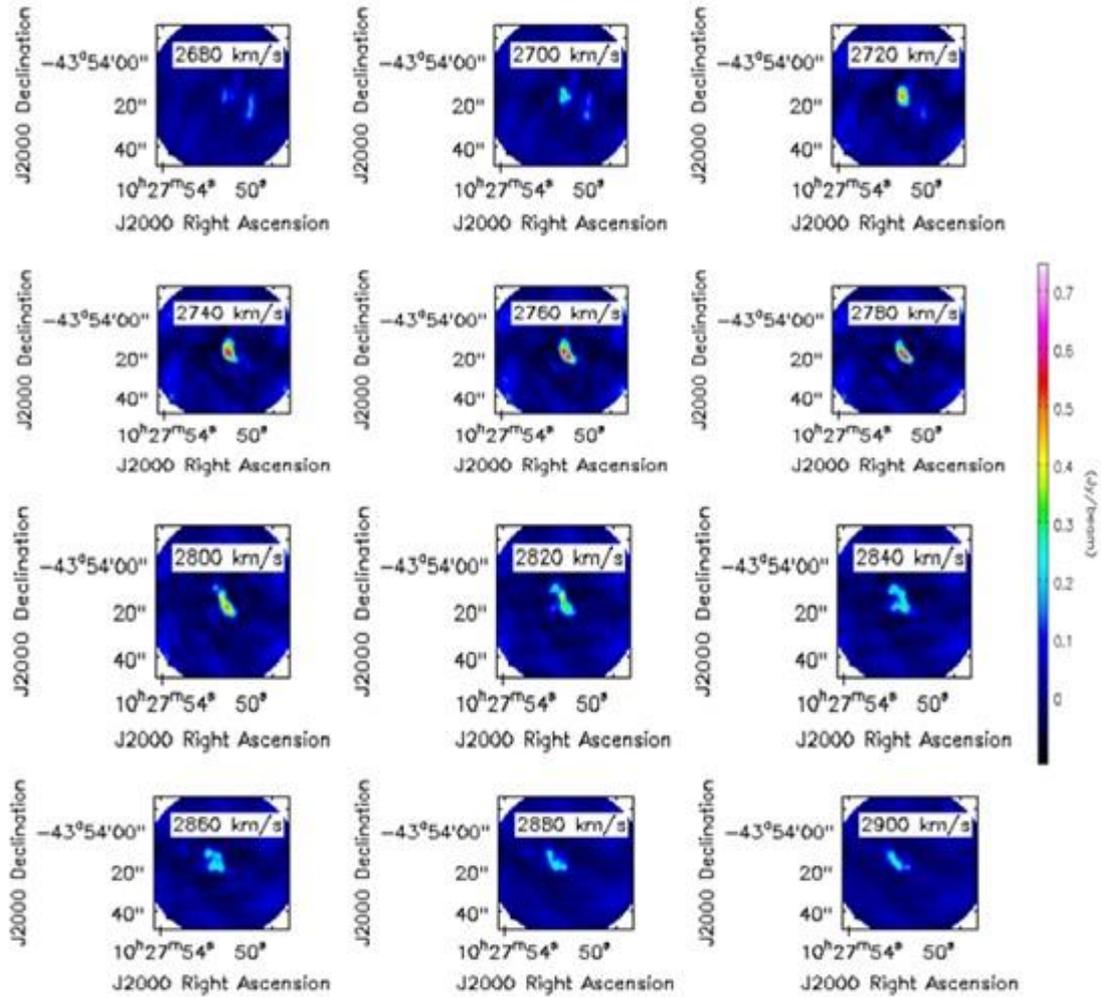

Figure 1. Channel maps of the CO(1-0) line emission in the central region of NGC 3256 with the velocity channel width= 20 km s$^{-1}$. The system velocity of NGC 3256 is 2775 km s$^{-1}$. The color scale range is shown in the wedge at right in units of Jy beam$^{-1}$.

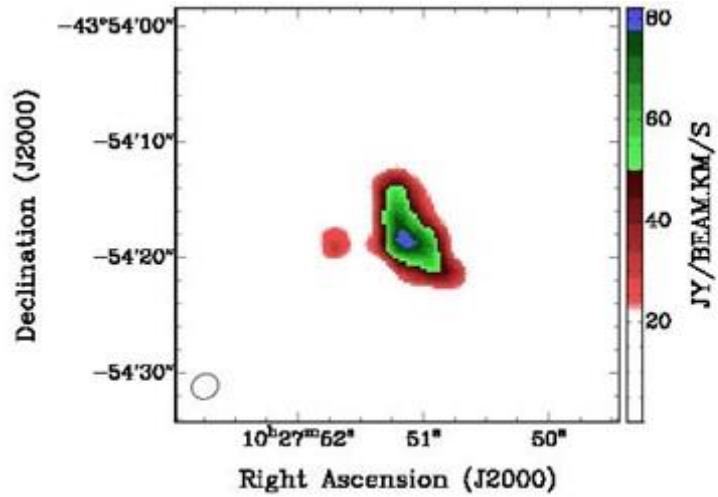

Figure 2. Integrated intensity (mom0) map of the CO(1-0) line emission in the central region of NGC 3256. The color scale range is shown in the wedge at right in units of Jy beam$^{-1}$ km s$^{-1}$. The synthesized beam is $2.3'' \times 2.10''$.

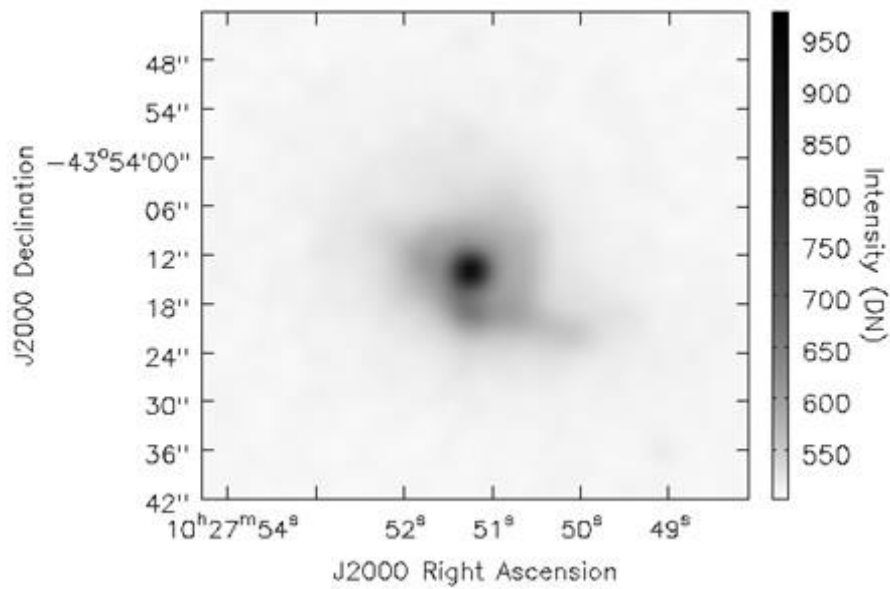

Figure 3. Image of NGC 3256 at the 2MASS Ks-band.

Table 1. Galaxy properties for NGC 3256.

| Parameter | Value |
|---|---|
| R.A.(J2000) | 10h27m51s.23 |
| Dec.(J2000) | $-43°54'16.6''$ |
| Adopted distance | 35 Mpc |
| Scale. 1" in pc | 170 |
| i | 30° |
| Systemic velocity | 2775 km s$^{-1}$ |
| Velocity resolution | 20 km s$^{-1}$ |
| z | 0.009354 |

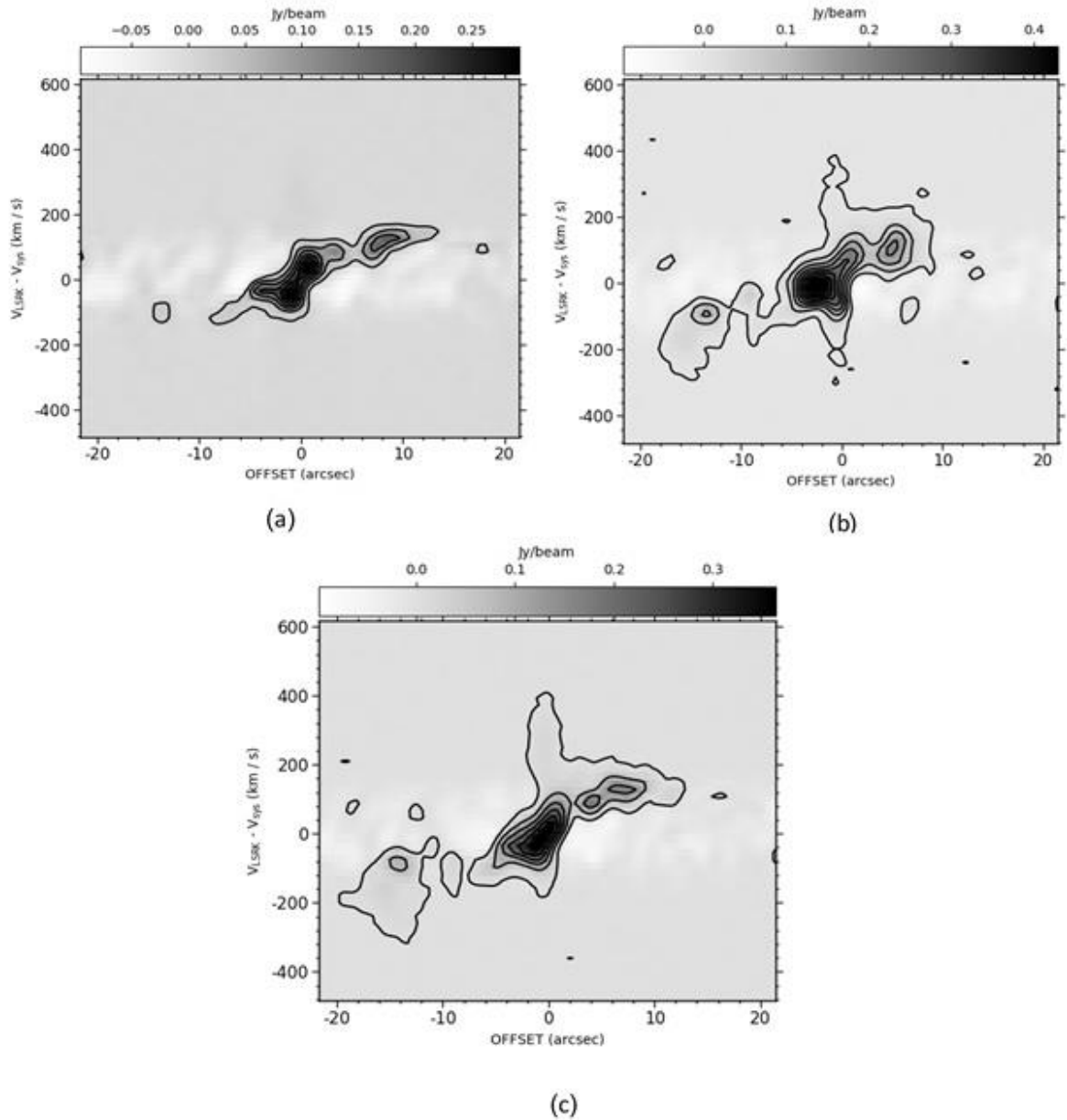

Figure 4. CO Position-Velocity diagrams of NGC 3256 passing through the nuclei along different PA angles. (a) PV diagram along PA= -90° through N nucleus, (b)PV diagram along PA= -90°through S nucleus, (c) PV diagram along PA= -90° passing through the center between the two nuclei. Each PV cut has a slit width of 1″.

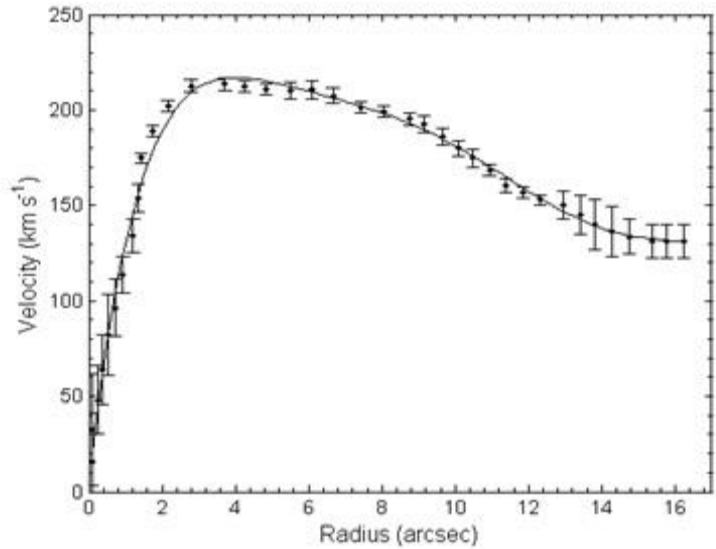

Figure 5. Projected rotation velocity of NGC 3256 derived from the PV diagram of Figure 4c assuming an inclination angle of 30°. Dots with error bars are the data points and the solid curve represents a 5th-order polynomial fitting to the data. The error bars are the standard deviations of the velocity.

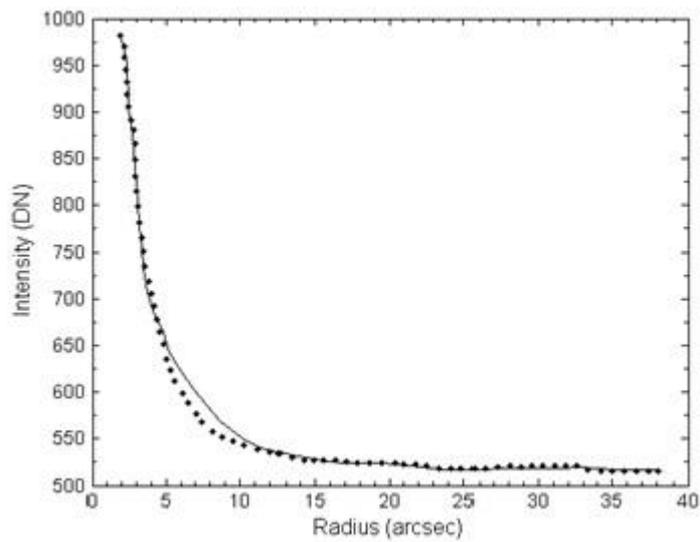

Figure 6. Surface brightness profile of NGC 3256 in 2MASS Ks image. The solid line is the radial profile of the Ks image of NGC 3256, and the dashed line is the fitting modified Hubble profile with $r_0 = 2.44''$ and $I_0 = 412.5$ data number (DN)

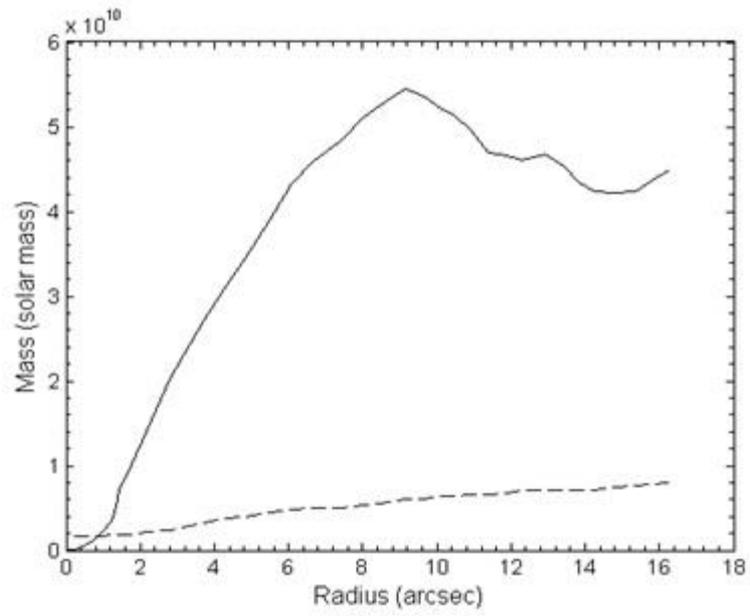

Figure 7. Total mass and baryonic mass of NGC 3256 as a function of radius. The solid line is the total mass of NGC 3256, and the dashed line is the baryonic component of NGC3256.